\documentclass[12pt,epsf]{article}
\usepackage{amsmath}
\usepackage{amssymb}
\begin{document}
\def\a{\alpha}
\def\b{\beta}
\def\c{\varepsilon}
\def\d{\delta}
\def\e{\epsilon}
\def\f{\phi}
\def\g{\gamma}
\def\h{\theta}
\def\k{\kappa}
\def\l{\lambda}
\def\m{\mu}
\def\n{\nu}
\def\p{\psi}
\def\q{\partial}
\def\r{\rho}
\def\s{\sigma}
\def\t{\tau}
\def\u{\upsilon}
\def\v{\varphi}
\def\w{\omega}
\def\x{\xi}
\def\y{\eta}
\def\z{\zeta}
\def\D{{\mit \Delta}}
\def\G{\Gamma}
\def\H{\Theta}
\def\L{\Lambda}
\def\F{\Phi}
\def\P{\Psi}

\def\S{\Sigma}

\def\o{\over}
\def\beq{\begin{eqnarray}}
\def\eeq{\end{eqnarray}}
\newcommand{\gsim}{ \mathop{}_{\textstyle \sim}^{\textstyle >} }
\newcommand{\lsim}{ \mathop{}_{\textstyle \sim}^{\textstyle <} }
\newcommand{\vev}[1]{ \left\langle {#1} \right\rangle }
\newcommand{\bra}[1]{ \langle {#1} | }
\newcommand{\ket}[1]{ | {#1} \rangle }
\newcommand{\EV}{ {\rm eV} }
\newcommand{\KEV}{ {\rm keV} }
\newcommand{\MEV}{ {\rm MeV} }
\newcommand{\GEV}{ {\rm GeV} }
\newcommand{\TEV}{ {\rm TeV} }
\def\diag{\mathop{\rm diag}\nolimits}
\def\Spin{\mathop{\rm Spin}}
\def\SO{\mathop{\rm SO}}
\def\O{\mathop{\rm O}}
\def\SU{\mathop{\rm SU}}
\def\U{\mathop{\rm U}}
\def\Sp{\mathop{\rm Sp}}
\def\SL{\mathop{\rm SL}}
\def\tr{\mathop{\rm tr}}

\def\IJMP{Int.~J.~Mod.~Phys. }
\def\MPL{Mod.~Phys.~Lett. }
\def\NP{Nucl.~Phys. }
\def\PL{Phys.~Lett. }
\def\PR{Phys.~Rev. }
\def\PRL{Phys.~Rev.~Lett. }
\def\PTP{Prog.~Theor.~Phys. }
\def\ZP{Z.~Phys. }

\newcommand{\drawsquare}[2]{\hbox{%
\rule{#2pt}{#1pt}\hskip-#2pt
\rule{#1pt}{#2pt}\hskip-#1pt
\rule[#1pt]{#1pt}{#2pt}}\rule[#1pt]{#2pt}{#2pt}\hskip-#2pt
\rule{#2pt}{#1pt}}

\def\vbr{\vphantom{\sqrt{F_e^i}}}
\newcommand{\fund}{\drawsquare{6.5}{0.4}}
\newcommand{\afund}{\overline{\fund}}
\newcommand{\symm}{\drawsquare{6.5}{0.4}\hskip-0.4pt%
        \drawsquare{6.5}{0.4}}
\newcommand{\asymm}{\raisebox{-3pt}{\drawsquare{6.5}{0.4}\hskip-6.9pt%
        \raisebox{6.5pt}{\drawsquare{6.5}{0.4}}}}
\newcommand{\asymmthree}{\raisebox{-7pt}{\drawsquare{6.5}{0.4}}\hskip-6.9pt%
\raisebox{-0.5pt}{\drawsquare{6.5}{0.4}}\hskip-6.9pt%
\raisebox{6pt}{\drawsquare{6.5}{0.4}}}
\newcommand{\asymmfour}{\raisebox{-10pt}{\drawsquare{6.5}{0.4}}\hskip-6.9pt%
\raisebox{-3.5pt}{\drawsquare{6.5}{0.4}}\hskip-6.9pt%
\raisebox{3pt}{\drawsquare{6.5}{0.4}}\hskip-6.9pt%
        \raisebox{9.5pt}{\drawsquare{6.5}{0.4}}}
\newcommand{\Ythrees}{\raisebox{-.5pt}{\drawsquare{6.5}{0.4}}\hskip-0.4pt%
          \raisebox{-.5pt}{\drawsquare{6.5}{0.4}}\hskip-0.4pt%
          \raisebox{-.5pt}{\drawsquare{6.5}{0.4}}}
\newcommand{\Yfours}{\raisebox{-.5pt}{\drawsquare{6.5}{0.4}}\hskip-0.4pt%
          \raisebox{-.5pt}{\drawsquare{6.5}{0.4}}\hskip-0.4pt%
          \raisebox{-.5pt}{\drawsquare{6.5}{0.4}}\hskip-0.4pt%
          \raisebox{-.5pt}{\drawsquare{6.5}{0.4}}}
\newcommand{\Ythreea}{\raisebox{-3.5pt}{\drawsquare{6.5}{0.4}}\hskip-6.9pt%
        \raisebox{3pt}{\drawsquare{6.5}{0.4}}\hskip-6.9pt
        \raisebox{9.5pt}{\drawsquare{6.5}{0.4}}}
\newcommand{\Yfoura}{\raisebox{-3.5pt}{\drawsquare{6.5}{0.4}}\hskip-6.9pt%
        \raisebox{3pt}{\drawsquare{6.5}{0.4}}\hskip-6.9pt
        \raisebox{9.5pt}{\drawsquare{6.5}{0.4}}\hskip-6.9pt
        \raisebox{16pt}{\drawsquare{6.5}{0.4}}}
\newcommand{\Yadjoint}{\raisebox{-3.5pt}{\drawsquare{6.5}{0.4}}\hskip-6.9pt%
        \raisebox{3pt}{\drawsquare{6.5}{0.4}}\hskip-0.4pt
        \raisebox{3pt}{\drawsquare{6.5}{0.4}}}
\newcommand{\Ysquare}{\raisebox{-3.5pt}{\drawsquare{6.5}{0.4}}\hskip-0.4pt%
        \raisebox{-3.5pt}{\drawsquare{6.5}{0.4}}\hskip-13.4pt%
        \raisebox{3pt}{\drawsquare{6.5}{0.4}}\hskip-0.4pt%
        \raisebox{3pt}{\drawsquare{6.5}{0.4}}}
\newcommand{\Yflavor}{\Yfund + \overline{\Yfund}} 
\newcommand{\Yoneoone}{\raisebox{-3.5pt}{\drawsquare{6.5}{0.4}}\hskip-6.9pt%
        \raisebox{3pt}{\drawsquare{6.5}{0.4}}\hskip-6.9pt%
        \raisebox{9.5pt}{\drawsquare{6.5}{0.4}}\hskip-0.4pt%
        \raisebox{9.5pt}{\drawsquare{6.5}{0.4}}}%

\baselineskip 0.7cm

\begin{titlepage}

\begin{flushright}
IPMU 08-0008\\
SLAC-PUB-13134\\
UCB-PTH-08/04
\end{flushright}

\vskip 1.35cm
\begin{center}
{\large \bf
Conformal Supersymmetry Breaking\\ and \\ Dynamical Tuning of the Cosmological Constant
}
\vskip 1.2cm
M. Ibe${}^{1}$, Y. Nakayama${}^{2}$ and T.T. Yanagida${}^{3,4}$

\vskip 0.4cm
${}^{1}$ {\it Stanford Linear Accelerator Center, Stanford University,
                Stanford, CA 94309 and} \\
{\it Physics Department, Stanford University, Stanford, CA 94305}\\
${}^{2}${\it Berkeley Center for Theoretical Physics and Department of Physics,
\\
University of California, Berkeley, California 94720-7300}

${}^3${\it Department of Physics, University of Tokyo,     Tokyo 113-0033, Japan}

${}^4${\it Institute for the Physics and Mathematics of the Universe, University of Tokyo, \\
Kashiwa 277-8568, Japan}

\vskip 1.5cm

\abstract{
We propose ``conformal supersymmetry breaking" models, which tightly relate the  conformal breaking scale (i.e. R-symmetry breaking scale) and the supersymmetry breaking scale. Both the scales are originated from the constant term in the superpotential through the common source of the R-symmetry breaking. We show that dynamical tuning between those mass scales significantly reduces the degree of fine-tuning necessary for generating the almost vanishing cosmological constant.}
\end{center}
\end{titlepage}

\setcounter{page}{2}

\section{Introduction}
The origin of mass scale in the standard model has always been a hot issue. Supersymmetry (SUSY) is an illuminating approach to this problem by avoiding otherwise uncontrollable quadratic divergences from radiative corrections. 
 In absence of the quadratic divergences, the mass scale of the standard model is tightly related to the SUSY breaking scale. The idea of dynamical SUSY breaking ~\cite{Witten:1981nf} is a next step to accomplish this pursuit by providing a naturally small SUSY breaking scale through dimensional transmutation. 

The success of the dynamical SUSY breaking, however, is still halfway.
In any theory with a spontaneous SUSY breaking, we need yet another dimensionful parameter in addition to the SUSY breaking scale: a constant term should be included in the superpotential in order to realize the flat universe after the SUSY breaking. Since the constant term is nothing to do with the SUSY breaking scale  in generic dynamical SUSY breaking models, the requirement of the flat universe leads to a severe fine-tuning.

In this letter, we try to answer this remaining half of the problem 
by introducing an idea of ``conformal SUSY breaking" based on ``conformal extension" of the dynamical SUSY breaking models. In this scenario, the conformal symmetry and its subsequent breaking at a lower-energy scale constrain the system in quite a non-trivial way. Almost all parameters of the theory are fixed by the conformal invariance. A remaining relevant deformation, typically a unique mass parameter of the model, then, disturbs the conformal invariance at the lower-energy scale. The resultant conformal symmetry breaking triggers the dynamical SUSY breaking, and consequently, the SUSY breaking scale is determined uniquely by the conformal breaking scale.

A far-reaching consequence of this scenario is that  when the conformal extended theory is strongly coupled, we can identify the origin of the unique mass parameter of the model with a constant term in the superpotential through the underlying R-symmetry breaking. This, in turn, gives us a dynamical tuning mechanism of the cosmological constant. The conformal SUSY breaking, therefore, consolidates two independent mass scales into one, completing our pursuit of its origin.

The organization of the paper is as follows. 
In section 2, we introduce conformally extended SUSY breaking models.
In section 3, we discuss the resolution of the cosmological constant problem through conformal SUSY breaking by identifying the constant term in the superpotential with a mass deformation of the conformally extended SUSY breaking models. In section 4, we show three 
explicit examples of the model.
The last section is devoted to our conclusions with a further discussion.

\section{Conformal SUSY breaking model}
A ``conformal SUSY breaking" is a novel way to understand the scale of the SUSY breaking. At first sight, as is clear from the definition, any characteristic energy scale cannot appear in conformal field theories. However, a key idea of the conformal SUSY breaking is to relate the conformal breaking scale (i.e. R-symmetry breaking scale\footnote{We further postulate that the R-symmetry breaking is originated from a constant term in the superpotential, $w_0$.}) and the SUSY breaking scale.\footnote{The superconformal symmetry automatically demands the existence of R-symmetry \cite{Dobrev:1985qv}. Therefore, the conformal symmetry breaking and the R-symmetry breaking is intimately related, which will be important later when we discuss the cosmological constant.} For this purpose, we introduce a concept of conformally extended dynamically SUSY breaking models.

The starting point is a particular class of dynamical SUSY breaking models which break SUSY from the strong gauge interaction. To construct their conformal extension, we add massive
$N_f$ vector-like representations ($P$, $\bar{P}$)  as new flavors. The mass is simply given by the superpotential:
\begin{eqnarray}
W = \sum m P \bar P \ . 
\end{eqnarray}
This extension is,  in a sense, trivial: the extended model reduces to
the original model after the new flavors decouple~\cite{Witten:1982df}, leading to a dynamical SUSY breaking at low energy.
Even worse, it might appear that the introduction of the new mass scale is a major drawback that spoils the understanding of the origin of the mass scale in the dynamical SUSY breaking model.

This extension, however, shows its real potential when $N_{f}$ is chosen in a specific range
so that the extended model has a non-trivial infrared (IR)-fixed point in the massless limit 
of the new flavors ($m\to 0$). As we will see shortly, the conformal dynamics essentially removes arbitrary dependence on the ratio between the dynamical scale and the newly added mass scale, avoiding the drawback just mentioned. 

Conformally extended SUSY breaking models have many examples. 
The extension of the non-calculable dynamical SUSY breaking model based on 
SO(10) gauge theory with a spinor representation~\cite{Affleck:1984mf,Murayama:1995ng}
 is known to have an non-trivial IR-fixed point 
for $7<N_{f}<21$~\cite{Pouliot:1996zh,Kawano:1996bd}.
As another example, the extension of the vector-like dynamical SUSY breaking model
based on the SP$(N)$ gauge theory~\cite{Izawa:1996pk,Intriligator:1996pu} 
also possesses an non-trivial IR-fixed point~\cite{Ibe:2005pj}.

To utilize the full power of conformal invariance, we make a crucial assumption that the conformally extended SUSY breaking model is in the vicinity of the IR-fixed point at the ultraviolet (UV) cut-off scale
where we can neglect the mass of the new flavors.
Under this assumption, all the coupling constants in the SUSY breaking sector immediately
converge to the values at the IR-fixed point once they evolve down to the IR from the UV cut-off scale.
Therefore, 
there remains no free parameter in the conformally extended SUSY breaking sector at the IR scale.

Down in the far IR limit, the conformal invariance is disturbed by the mass term of the new flavors. Since the newly added flavor decouples at the scale
\begin{eqnarray}
\label{eq:Pmass}
m_{\mathrm{phys}} = m \left(\frac{m}{M_{\rm UV}}\right)^{\frac{\gamma_P}{1-\gamma_P}} \ ,
\end{eqnarray}
the SUSY is dynamically broken below the scale. Here $\gamma_{P}$ denotes the anomalous dimension of $\bar{P} P$ at the IR-fixed point, and
$M_{\rm UV}$ is the scale of the UV cut-off.
Notice that since all the coupling constants of the SUSY breaking sector are fixed at the IR-fixed point, the SUSY breaking scale is related to the conformal breaking scale induced by the mass of the new flavors in a unique fashion:
\begin{eqnarray}
\label{eq:decouple}
\Lambda_{\rm susy} \simeq c_{\rm susy}m_{\rm phys} 
\end{eqnarray}
with a coefficient $c_{\rm susy}$. We emphasize that the ratio $c_{\rm susy}$ is not a free parameter of the model but determined by the dynamics.
When the model is strongly coupled at the IR-fixed point, $c_{\rm susy}$ is expected to be $O(1)$ because the gauge coupling constant of the SUSY breaking sector blows up just below the decoupling scale of  the new flavors.

To sum up, in the conformal SUSY breaking scenario, 
the SUSY breaking scale is uniquely determined by the physical mass of the new flavors, $m_{\rm phys}$.%
\footnote{In some dynamical SUSY breaking models, the conformal extension has extra benefits: the conformal dynamics fixes all the dimensionless parameters at the IR-fixed point value, reducing the number of otherwise freely-tunable parameters in the original model and increasing the predictability.}
The physical mass, $m_{\rm phys}$, depends on the two mass parameters, $m$ and $M_{\rm UV}$.
In the following, we also assume that the UV cut-off scale is close to the reduced Planck scale 
$M_{\rm PL} \simeq 2.4 \times 10^{18}$\,GeV, $M_{\rm UV}\simeq M_{\rm PL}$, 
which is naturally realized if the coupling constants are close 
to the value at the IR-fixed at the Planck scale.


\section{Dynamical tuning of the cosmological constant}
Now, let us move on to the remaining half of the problem, i.e. the relation between
the SUSY breaking scale and the scale of the constant term in the superpotential.
In the course of the discussion, we will see the relevance of the strongly coupled conformal 
dynamics to the cosmological constant problem.

Whenever SUSY is broken, a positive contribution to the cosmological constant from the SUSY breaking source, $\Lambda_{\rm susy}^{4}$, 
should be cancelled by a constant term in the superpotential 
to obtain the observed (almost) flat universe:
\begin{eqnarray}
\label{eq:flat}
\vev{V} = \Lambda_{\rm susy}^{4} - 3 \frac{|w_{0}|^{2}}{M_{\rm PL}^{2}} = \Lambda_{\rm c.c.}^{4}
\simeq 0 \ .
\end{eqnarray}
Here, $\Lambda_{\rm c.c.}^{4}$ is the cosmological constant,
and $w_{0}$ is the constant term in the superpotential:
\begin{eqnarray}
W = w_{0}\ . 
\end{eqnarray}
Thus, even if we explained the SUSY breaking scale by dimensional transmutation of the strong dynamics, we still would need to explain the origin of the constant term.
Furthermore, in terms of the degree of fine-tuning for the cosmological constant problem, since the constant term $w_{0}$ is independent of the SUSY breaking scale, the usual  models require $O(\Lambda_{c,c}^{4}/M_{\rm PL}^{4})$
fine-tuning.

%
The above expression of the cosmological constant in Eq.~(\ref{eq:flat}), however, is showing an important hint to reduce the cosmological constant problem. 
The key is that the both terms in Eq.~(\ref{eq:flat}) represent two different symmetries, i.e. the supersymmetry and the R-symmetry.
This fact suggests that the cancellation between these two terms might
be naturally understood in a theory where the R-symmetry breaking and the supersymmetry
breaking are related with each other.
Fortunately, the conformal SUSY breaking model developed in the last section provides
such a tight link between the R-symmetry and the supersymmetry breakings,
where both symmetries are parts of the superconformal symmetry.
There, the explicit R-symmetry breaking in the superconformal algebra, (i.e. the mass term of $P$),
triggers the dynamical supersymmetry breaking.
Therefore, once we find an appropriate relation between the R-breaking mass term of $P$, $\bar{P}$
and the R-symmetry breaking which shows up in Eq.~(\ref{eq:flat}), 
we can link the two terms in Eq.~(\ref{eq:flat}) together and might reduce the cosmological constant 
problem.

As the simplest proposal for such a relation between two sources of the R-symmetry breaking,
we consider that the two breakings have the same origin, the constant term in the superpotential.
That is, we assume the mass term of $P$ and $\bar P$ is given by,
\begin{eqnarray}
\label{eq:mass}
W = w_{0} \left( 1 + \frac{c_{m}}{M_{\rm PL}^{2}} P\bar{P} \right) \ 
\end{eqnarray}
with a coefficient $c_{m}=O(1)$ (i.e. $m= c_{m}w_{0}/M_{\rm PL}^{2}$)
at the Planck scale $M_{\rm PL}$. 
This linear identification of the constant term and the mass term may be understood
naturally if one conceives that those terms are controlled by a classical R-symmetry under 
which new flavors $P$ and $\bar{P}$ have 0 charges,%
\footnote{In many cases (e.g. in conformally extended SO(10) model in section 4),
this classical R-symmetry is anomaly-free, and hence,
appears as a quantum R-symmetry even at the interacting fixed point.} 
which is different from the conformal R-symmetry above on the IR-fixed point.


Under this assumption, we further suppose that 
the conformally extended model is strongly interacting at the IR-fixed point 
and the anomalous dimension of the new flavors is close to the unitarity bound 
$\gamma_{P}\simeq -1$~\cite{Mack:1975je}. 
Then, new flavors decouple at the scale
\begin{eqnarray}
\label{eq:Pmass2}
m_{\rm phys} \simeq \sqrt{c_{m} \frac{w_{0}}{M_{\rm PL}}} \ ,
\end{eqnarray}
where we have used Eq.~(\ref{eq:Pmass}) with $\gamma_{P}\simeq -1$.
By recalling that the scale of the dynamical SUSY breaking 
in the strongly coupled conformally extended model 
is given by the decoupling scale of the new flavors (Eq.~(\ref{eq:decouple})),
the above mass of the new flavors results in
\begin{eqnarray}
\label{eq:susy}
\Lambda_{\rm susy}\simeq c_{\rm susy} \sqrt{c_{m} \frac{w_{0}}{M_{\rm PL}}}\ 
\end{eqnarray}
with $c_{\rm susy} = O(1)$.
Remarkably, the resultant SUSY breaking scale is the mass scale just required to realize the flat universe.
In fact, by plugging Eq.~(\ref{eq:susy}) into Eq.~(\ref{eq:flat}),
the flat universe condition is satisfied for $c_{m}=O(1)$.%
\footnote{
A similar tuning  of the cosmological constant works in 
the meta-stable supersymmetry breaking model of Intriligator-Seiberg-Shih 
(ISS)~\cite{Intriligator:2006dd} 
without the conformal dynamics if the dynamical scale $\Lambda_{\rm dyn}$ 
in the electric theory is of the order of the Planck scale, $\Lambda_{\rm dyn}\simeq M_{\rm PL}$. 
Suppose the quark mass $m$ is also given by Eq.~(\ref{eq:mass}). 
Then, the SUSY breaking scale in the ISS model is
$\Lambda_{\rm SUSY} = \sqrt{m\Lambda_{\rm dyn}}\simeq \sqrt{m\rm M_{PL}}\simeq 
\sqrt{c_mw_0/M_{\rm PL}}$,
which leads to the desired cancellation as shown in the text.}

Our scenario is summarized as follows.
The mass term of the new flavors given in Eq.~(\ref{eq:mass})
grows up as it evolves down to the IR scale, eventually breaking the conformal symmetry to trigger the SUSY breaking.
In this process, the resultant SUSY breaking scale $\Lambda_{\rm susy}$ is dynamically tuned to 
the  scale which is appropriate for the cancellation of the cosmological constant from the negative contribution of the constant term in the superpotential. The flat universe condition is assisted by the  ``dynamical tuning'' of the SUSY breaking scale, and the degree of fine-tuning is significantly 
reduced from 
$O(\Lambda_{c.c.}^{4}/M_{\rm PL}^{4})$
to
$O(\Lambda_{c.c.}^{4}/\Lambda_{\rm susy}^{4})$.%
\footnote{
If one assumes the landscape of vacua and considers that 
one of  the two contributions in Eq.~(\ref{eq:flat}) obeys the logarithmic distribution,
the degree of fine-tuning can be also reduced to 
$O(\Lambda_{c.c.}^{4}/\Lambda_{\rm susy}^{4})$
without any direct link between the $R$-symmetry 
and the supersymmetry breakings (see related discussions in Refs.~\cite{Dine:2004is,Dine:2005gz}).
In any known approach including ours, however, 
it seems very difficult  to reduce the problem any further.}

To emphasize  the importance of the strong conformal dynamics,
we attempt to achieve the dynamical tuning of the cosmological constant
for a generic value of $\gamma_{P}$. 
Again, by assuming the mass term of the new flavors in Eq.~(\ref{eq:mass}),
we obtain a required value of $c_{m}$ which satisfies the flat universe condition Eq.~(\ref{eq:flat})
as
\begin{eqnarray}
c_{m} =  (3 c_{\rm susy}^{-2})^{(1-\gamma_{P})/2}
\left( \frac{M_{\rm PL}^{3}}{w_{0}} \right)^{(1+ \gamma_{P})/2} \ .
\end{eqnarray}
Therefore, the required value of $c_{m}$ is of the $O(1)$
for  $\gamma_{P}\simeq -1$ and $c_{\rm susy}=O(1)$, while it
gets larger when $\gamma_{P}$ becomes larger than $-1$.

It is also interesting to note that for a generic value of $\gamma_{P}$, the gravitino mass is given by
\begin{eqnarray}
m_{3/2} = \frac{w_0}{M_{\rm PL}^2} = M_{\rm PL} (3c^{-2}_{\rm susy})^{\frac{1-\gamma_P}{1+\gamma_P}} c_m^{-2/(1+\gamma_p)} \ .
\end{eqnarray}
Thus, for $\gamma_P \simeq -1+\epsilon$ with $\epsilon >0$ (because of unitarity), 
we may be able to understand the huge hierarchy between the Planck scale and the SUSY breaking scale through the enhancement of the small number $c_m$ and the cancellation of the cosmological constant.

\section{Examples of strongly coupled conformal extension}
We present three examples of the strongly coupled conformal
extension of the dynamical SUSY breaking which realize the above mechanism.%
\footnote{
In a model proposed in Ref.~\cite{Kitano:2006wm}, the SUSY breaking scale and
the constant term in the superpotential is tuned by a similar dynamics.
However, since that model introduces an extra mass scale that has nothing to do with the R-symmetry breaking constant $w_{0}$, it does not fit into the realm of our conformal SUSY breaking models.} 
We will briefly mention phenomenological applications of these models in the next section.

The first example is the conformal extension of the dynamical SUSY breaking model of SO$(10)$
gauge theory with a spinor representation~\cite{Affleck:1984mf,Murayama:1995ng}
 augmented by $N_{f}$ flavors ($7<N_{f}<21$) in vector representation.
As analyzed in Refs.~\cite{Ibe:2005qv,Kawano:2005nc}
the anomalous dimensions of the chiral superfields at the conformal fixed point can be computed by using the $a$-maximization technique~\cite{Intriligator:2003jj},
and that for the added flavors is given by  $\gamma_{P}\simeq -0.97$ for $N_{f} = 10$.
Therefore, the conformal extension of the SO$(10)$ dynamical SUSY breaking model is a good example of the conformal SUSY breaking model we have proposed in this letter.

The second example is a conformal extension of the SP$(N_{c})$ vector-like SUSY breaking 
model~\cite{Izawa:1996pk,Intriligator:1996pu} which consists of 
$2(N_{c}+1)$ fundamental representations and $(N_{c}+1)(2N_{c}+1)$ singlets.
With an appropriate number of additional flavors, the model can be extended to the 
conformal SUSY breaking model~\cite{Ibe:2005pj}, and the anomalous dimensions 
are again determined by the $a$-maximization technique.
Among this class of the conformal SUSY breaking model, the model based on SP(2) 
with two additional massive flavors (or 4 additional fundamental representations) 
is a good instance of our proposal in which the anomalous dimension of the new flavors 
is given by $\gamma_{P}\simeq -0.98$.

The third example is a model based on SP$(3)\times$SP(1)$^{2}$ gauge theory~\cite{Ibe:2005pj} 
which is also a conformal extension of SP$(3)$ vector-like SUSY breaking 
model~\cite{Izawa:1996pk,Intriligator:1996pu} with added bi-fundamental matter $P, \bar{P}$.
As discussed in Ref.~\cite{Ibe:2005pj}, the anomalous dimension of the new flavors are given by
$\gamma_{P}=-1$.
Thus, this model is also a good example of our proposal.

\section{Conclusion and Discussion}
In this letter, we have proposed a new concept of conformal SUSY breaking, 
which makes it possible to relate the conformal breaking scale (i.e. R-symmetry breaking scale) and the SUSY breaking scale. In the strongly coupled case, we can further identify the origin of the SUSY breaking scale with the constant term in the superpotential thorough the underlying R-symmetry breaking.
As a result, we found that the degree of fine-tuning for the cosmological constant problem
can be tremendously reduced by the dynamical tuning between those parameters. 

So far, we have concentrated on the mass scales of the SUSY breaking sector.
The mass scale of the SUSY standard model (SSM) sector, on the other hand, depends on how the SUSY breaking effects are mediated.

For example, if the breaking effects are mediated by the supergravity effects, the mass scale
of the SSM is given by $\Lambda_{\rm susy}^2/M_{\rm PL}$ which is also controlled by
the constant term $w_{0}$ via Eq.~(\ref{eq:susy}).
A concrete construction of the gravity mediation~\cite{Nilles:1983ge} in the conformal SUSY 
breaking model is realized by identifying the operator $\bar{P}P$ with the effective
Polonyi field $S$, i.e. $S\sim \bar{P}P$ which is expected to develop a non-vanishing $F$-term 
condensation at the SUSY breaking scale
\begin{eqnarray}
 \vev{\bar{P}P}|_{F-{\rm term}} \sim \Lambda_{\rm susy}^{2}\theta^{2}.
\end{eqnarray}
Under this identification, the gaugino mass, for example, is given by the higher dimensional 
operator in the superpotential at the Planck scale,
\begin{eqnarray}
 W = \frac{\bar{P}P}{M_{\rm PL}^{2}} {\cal W}{\cal W},
\end{eqnarray}
where ${\cal W}$ denotes the field strength chiral supermultiplet of the SSM gauge group.
By the SUSY breaking scale, the above higher dimensional operator has been 
enhanced by the same factor of the mass term of $P$ and $\bar{P}$ given in Eq.~(\ref{eq:Pmass})
and gives the SMM gaugino mass of the order of $m_{3/2}$ when the anomalous dimension 
$\gamma_{P}$ is close to $-1$.
The $\mu$ and $B\mu$ term are also provided via the 
so called Giudice-Masiero mechanism in terms of $S\sim\bar{P}P$~\cite{Giudice:1988yz}
which are again enhanced by the same enhancement factor in Eq.~({\ref{eq:Pmass}}).

The advantage of the above construction of the gravity mediation in the conformal
SUSY breaking is the absence of the complete neutral chiral superfield which
causes the cosmological problems known as the Polonyi problem~\cite{Coughlan:1983ci} 
and  the Polonyi induced gravitino problem~\cite{Ibe:2006am}.
In our construction, the origin of the effective Polonyi field has a definite meaning, 
and it is expected to be preferred by the large curvature effects in the early universe.
Therefore, we can solve the Polonyi/Polonyi-gravitino problems which is
caused by the large amplitude of the coherent oscillation of the Polonyi field after inflation
(see earlier attempts to the solution to the Polonyi problem in Ref.~\cite{Dine:1983ys}.).


As for the gauge mediation models~\cite{Giudice:1998bp},
the situation is much more complicated since they require another mass scale as the messenger scale.
It is, however, possible to construct concrete models where the mass scale of the SSM
is related to only one mass parameter, the constant term $w_{0}$. 
For example, let us apply the second example of the conformal SUSY breaking in section~4
to the model with gauge mediation discussed in Refs.~\cite{Dine:1994vc,Fujii:2003iw}.
This example possesses a SU(2) flavor symmetry of the additional two flavors.
Now, let us use the flavor symmetry as a gauge symmetry and introduce a pair of SU(2) doublets, $E$ and $\bar E$, and a SU(2) singlet, $S$.%
\footnote{$E$, $\bar E$ and $S$ are neutral under SP(2).}
They couple with each other in the superpotential;
\begin{eqnarray}
\label{eq:gm}
W = f S E \bar E + \frac{\lambda}{3} S^{3} +  k S \psi \bar\psi,
\end{eqnarray}
where $f$, $\lambda$ and $k$ are dimensionless coupling constants.
We have also introduced messenger particles $\psi$  and $\bar \psi$
which are charged under the SSM gauge groups.
As discussed in Refs.~\cite{Dine:1994vc,Fujii:2003iw}, 
once $E$ and $\bar E$ obtain a positive SUSY breaking mass squared through the SU(2) 
interaction,%
\footnote{The sign of the mass squared of $E$ and $\bar E$ is affected by the strong interaction
of the SP(2), and they are not calculable~\cite{Ibe:2007wp}. } 
$S$ obtains a negative SUSY breaking mass squared by the effect of the first interaction
in Eq.~(\ref{eq:gm}).
The negative mass squared, then, results in non-vanishing vacuum expectation values (VEVs)
of the scalar and $F$-term of $S$ which result in 
a supersymmetric mass and a SUSY breaking mass of the messenger fields, respectively. 
As a result, we obtain the masses of the SSM fields after the messenger
fields integrated out.
In this construction, we have not introduced any dimensionful parameters
besides $w_{0}$.
Therefore, we can construct a model where the masses of the SSM fields 
are proportional to the constant term in the superpotential, $w_{0}$.

\section*{Acknowledgments}
The work of MI was supported by the U.S. Department of Energy under contract number 
DE-AC02-76SF00515.
The research of Y.~N. is supported in part by NSF grant PHY-0555662 and the UC Berkeley Center for Theoretical Physics.
This work was supported by World Premier International Research Center
Initiative (WPI Program), MEXT, Japan.

\end{document}